\documentclass[superscriptaddress,amsmath,amssymb,
aps,prl,longbibliography,twocolumn]{revtex4-2}

\usepackage{mathrsfs}
\usepackage{amssymb}
\usepackage{amsmath}
\usepackage{graphicx}% Include figure files showpacs
\usepackage{subfigure}
\usepackage{dcolumn}% Align table columns on decimal point
\usepackage{url}
\usepackage{color}
\usepackage[colorlinks=true,
            linkcolor=red,
            citecolor=blue,
            urlcolor=blue]{hyperref}
\usepackage{paralist}
\usepackage{ulem}
\usepackage{soul}

\newcommand{\sgn}{\mathrm{sgn}}

\begin{document}

%\preprint{APS/123-QED}

\title{The  Lieb excitations and topological flat mode of  spectral function of Tonks-Girardeau gas in Kronig-Penney potential}

\author{Wen-Bin He}
\email{wenbin.he@oist.jp}
\affiliation{Quantum Systems Unit, Okinawa Institute of Science and Technology Graduate University, 904-0495 Okinawa, Japan}%
 
\author{Giedrius \v{Z}labys}%
%\email{giedrius.zlabys@oist.jp}
\affiliation{Quantum Systems Unit, Okinawa Institute of Science and Technology Graduate University, 904-0495 Okinawa, Japan}%

\author{Hoshu Hiyane}%
%\email{Hoshu.Hiyane@oist.jp}
\affiliation{Quantum Systems Unit, Okinawa Institute of Science and Technology Graduate University, 904-0495 Okinawa, Japan}%

\author{Sarika Sasidharan Nair }%
%\email{sarika.sasidharan@oist.jp}
\affiliation{Quantum Systems Unit, Okinawa Institute of Science and Technology Graduate University, 904-0495 Okinawa, Japan}%

\author{Thomas Busch}%
\email{thomas.busch@oist.jp}
\affiliation{Quantum Systems Unit, Okinawa Institute of Science and Technology Graduate University, 904-0495 Okinawa, Japan}%

\date{\today}

\begin{abstract}
    Lieb excitations are fundamental to the understanding of the low energy behaviour of many-body quantum gases. Here we study the spectral function of a Tonks-Girardeau gas in a finite sized Kronig-Penney potential and show that the Lieb-I and Lieb-II excitations can become gapped as a function of the barrier height. Moreover, we reveal the existence of a topological flat mode near the Fermi energy and at zero momentum and show that this is robust to perturbations in the system. Through a scaling analysis, we determine the divergent behaviour of the spectral function. Our results  provide a significant reference for the observation and understanding of the gapped Lieb excitations and the topological flat mode of quantum gases in experimentally realistic subwavelength optical lattice potentials.
\end{abstract} 

\maketitle

\section{Introduction}
The spectral function describes the elementary excitations (EE) of a system in frequency-momentum space and is a key tool for understanding the low-energy properties of quantum many-body systems.
For condensed matter systems this quantity can be readily measured using angle-resolved photon emission spectroscopy (ARPES), which allows one to obtain detailed information about the band structure of a material \cite{zxshen}. 
For synthetic quantum many-body systems, such as ultracold atoms in optical lattices~\cite{Bloch_rmp}, different techniques such as photoemission or Bragg spectroscopy have been used to study, for example, the single-particle excitation spectrum of a Fermi gas  \cite{JinDS} or detection of spin-charge excitations \cite{Hulet2022}.
Because of the flexibility that ultracold atomic systems offer \cite{kinoshita2006,Jiang_2015,yang2017,minguzzi2022} they play an increasingly important role in understanding the elementary excitations in complex quantum many-body systems \cite{Hulet2022,Settino,cheng2022exact}.

Low-dimensional systems often exhibit different phenomena compared to three-dimensional systems. 
For instance, while in three-dimensional Bose-Einstein condensates (BECs) a gapless Bogoliubov mode emerges as an elementary excitation due to the spontaneous $U(1)$-symmetry breaking, the absence of phase coherence in one-dimension leads to a different excitation dynamics. More specifically, low-energy excitation spectrum of the one-dimensional Bose gas with contact-type interaction (the so-called Lieb--Liniger model), which is exactly solvable by means of the Bethe ansatz \cite{Bethe},  possesses two branches: 
the first one is the so-called Lieb-I branch that describes particle-like excitations, while the second branch, called the  Lieb-II, corresponds to hole-like excitations~\cite{Lieb1,Lieb2,caux2007jsm,Glazman2007,Glazman2008}. 

It is interesting to ask the question of how these well understood excitation behaviours change when the external geometry changes. One famous example here are systems which can support non-trivial topological states, which ever since the discovery of the Quantum Hall Effect \cite{Klitzing} have been recognised as playing a significant role in the quantum mechanical behaviour of condensed matter systems. More recently, topological properties have also been studied in ultracold atomic gases \cite{Cooper}, quantum optical systems \cite{Ozawa}, and have been considered for applications in quantum computation \cite{Nayak}. Non-trivial topological systems can be characterized by a topological invariant, for example the  Chern number \cite{TKNN} or the winding number \cite{ssh}, and the appearance of edge states is a typical property of such systems. These states are unusually robust states with wave functions localised at the edge of the system, and have become significant elements in the description of quantum materials. The influence of topological edge states on the elementary excitation of quantum many-body systems, especially the Lieb-excitations, is therefore an interesting problem that we consider in the following.

In this work we study the spectral function of a one-dimensional ultracold Bose gas in the strongly correlated Tonks-Girardeau (TG) limit. 
We assume that this system is placed in a finite sized Kronig-Penney potential \cite{Reshodko_2019} and utilize the exact solution of the many-body wave function to calculate the spectral function \citep{Settino}. We show that a topological flat mode appears in the spectrum, which is controlled by the location of the barrier potential with respect to the edge of the system.
Furthermore, the periodic barrier potential also breaks the symmetry of particle-hole excitations and leads to the two types of Lieb excitations becoming gapped at zero momentum, with the size of the gap being directly related to the height of barrier potentials.
Our results connect the understanding of topological edge modes in single particle systems with the excitations in many-body systems and extend the understanding of the Lieb-excitations in this strongly correlated system. 
In the following we  will first introduce the finite size Kronig-Penney model and the method to calculate the spectral function of the many-body TG gas. We then discuss the spectral function in detail, focusing on the gap between the two branches of excitations, and the flat mode that appears as a function of the system parameters. We will also study the form of the Fermi edge singularity of two Lieb excitations of TG gas, and check the robustness of the many-body edge mode against random perturbations of the barrier potentials.

\begin{figure}[tb]
    \includegraphics[width=\linewidth]{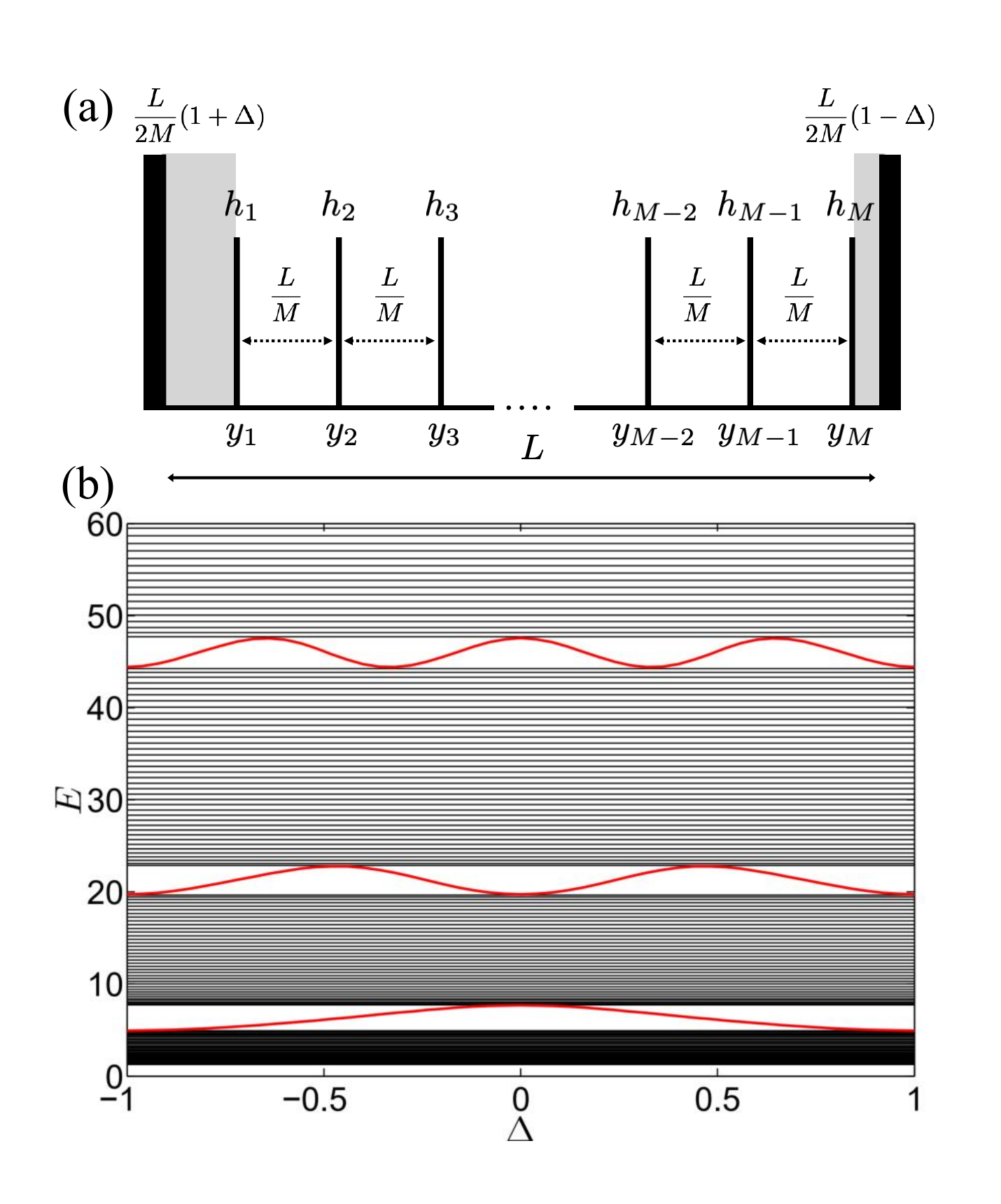} 
    \caption{(a). Schematic of the potential of the Arbitrary Finite Kronig-Penney model for a given shift parameter $\Delta$, which determines the position of the rigid barrier potential within the box (see text). (b) The single particle spectrum as a function of $\Delta$ for a system with parameters $L=40, M=40$ and $h=1.6$. The edge states exist inside the bands and are indicated in red.}
    \label{fig:Schematic}
    %, basis $N_{c}=256$
\end{figure}

\section{AFKP model and TG gas}

The system we  consider is an effectively one-dimensional, ultracold Bose gas trapped in a variant of the seminal Kronig-Penney (KP) potential \cite{kp1931}. Even though in real condensed matter systems the concept of point-like barriers is only a crude approximation, in ultracold atom systems point-like nanoscale potentials were recently shown to be realisable theoretically and experimentally \cite{Zoller2016,Kruckenhauser}. 
The simplest finite size KP model then consists of a finite sized box, in which a finite number of point-like barriers of arbitrary height are placed at arbitrary positions. For this the single particle Hamiltonian inside the box potential can be written as
\begin{equation}
    H=-\frac{\hbar^2}{2m}\frac{\partial^2}{\partial x^2}+\sum_{i=1}^{M} h_{i} \delta (x-y_{i}),
\label{Hkp}
\end{equation}
where $M$ is the number of barriers, the $\{ y_{i}\}$ are their positions and the $\{ h_{i}\}$ are their heights. The position coordinates $x$ and $\{ y_{i}\}$ are within the box, which ranges from $[-L/2,L/2]$. This model has been referred to as the Arbitrary Finite Kronig-Penney (AFKP) model, and despite the large number of free parameters that it allows for, its single particle spectrum can still be explicitly found using the Bethe ansatz \cite{Reshodko_2019}.

For ease of notation we will in the following use reduced dimensions by setting $\frac{\hbar^2}{m}=1$. We will also restrict ourselves to a lower number of free parameters by assuming that all barriers have the same height ($h_i=h$), and the distance between neighbouring barriers is constant. 
This still leaves one interesting free parameter, which corresponds to the placement of the (now rigid) lattice inside the finite sized box and we parametrise the position of the lattice as $y_{i}=-\frac{L}{2}+(i+\frac{\Delta-1}{2})\frac{L}{M}$  (see Fig.~\ref{fig:Schematic}(a)). Here $\Delta\in \left[-1,1 \right]$ is a shift parameter and for $\Delta=-1$ the leftmost barrier overlaps with the left hand side wall of the box, whereas for $\Delta=1$ the rightmost barrier overlaps with the right hand side wall of the box.

While the Bethe ansatz solution for the AFKP allows one in principle to also find the eigenstates of the system \cite{Reshodko_2019}, the numerical effort for solving the implicit Bethe equations is similar to straightforwardly diagonalising the full Hamiltonian numerically. For this we write the AFKP Hamiltonian in the basis of the wave functions of the infinite well 
\begin{equation}
    \varphi_{n}=\sqrt{\frac{2}{L}}\sin\left(k_{n}\left(x-x_{c}+\frac{L}{2}\right)\right),
\label{eq:phi_n}
\end{equation}
where the momenta are given by $k_{n}=\pi n/L$ with $n=1,2,3,\dots$, and $x_{c}$ is the center of the box size. With these the matrix elements of the AFKP Hamiltonian can be written as
\begin{align}
    H_{mn}&= \langle \varphi_{m} \vert H \vert \varphi_{n} \rangle\nonumber\\
    &=\langle \varphi_{m} \vert \hat{T} \vert \varphi_{n} \rangle +\langle \varphi_{m} \vert \hat{V} \vert \varphi_{n} \rangle=T_{mn}+V_{mn},
\end{align}
with
\begin{align}
    T_{mn}&=\frac{k^2}{2} \delta_{mn},\\
    V_{mn}&=\sum_{i=1}^{M} \frac{2h_{i}}{L} \sin\left(k_{m}\left(y_{i}-x_{c}+\frac{L}{2}\right)\right)\nonumber\\
    &\qquad\qquad\qquad\times\sin\left(k_{n}\left(y_{i}-x_{c}+\frac{L}{2}\right)\right).
\end{align}
The resulting matrix of $H$ can then be diagonalised  numerically to obtain the energy spectrum $\{ E_{n} \}$ and the single particle eigenfunctions $\{ \phi_{n}(x) \}$.  Having access to the latter is important for calculating the spectral function of the many-body systems in the strongly correlated TG regime. Let us mention that in this numerical approach it is crucial to choose a cut-off in momentum space $N_{c}$ that is large enough to ensure convergence for all physical quantities, and all results in the work have been checked for convergence.  Compared to using a finite difference method, the diagonalisation in momentum space can reduce  numerical error influencing the wavefunctions obtained.

The spectrum of the AFKP model as a function of the shift parameter $\Delta$ is shown in Fig.~\ref{fig:Schematic}(b) and consists of energy bands made up of $M$ levels each. Within each band the lower $(M-1)$  levels can be seen to be independent of the shift parameter, while the $M^\text{th}$ level is located in the subsequent gap and corresponds to a topological edge state (see the red lines in Fig.~\ref{fig:Schematic}(b)) \cite{Reshodko_2019}. These edge states are chiral as a function of $\Delta$ and are therefore located at the different edges of the box \cite{Reshodko_2019}. 

While the AFKP model has no closed form solutions for a gas of particles with arbitrary interaction strengths, it is possible to find the solution in the limit of infinitely strong interactions (the so-called Tonks-Girardeau limit) based purely on the knowledge of the single particle states.
This is due to the so-called Bose-Fermi mapping theorem, which allows one to describe a strongly interacting bosonic gas of $N$-particles by first constructing the many-body solution for a corresponding non-interacting Fermi gas, $\Psi_F$, and then symmetrizing it to go back to the bosonic solution \cite{Girardeau}
\begin{equation}
    \Psi_{B}(x_{1},\dots,x_{N})=A(x_{1},\dots,x_{N})\Psi_{F}(x_{1},\dots,x_{N}).
\end{equation}
Here, the unit antisymmetric function is given by $A(x_{1},\cdots,x_{N})=\prod_{1 \leq i<j\leq N} \sgn(x_{i}-x_{j})$.
The many-body Fermi wave function can be constructed as the Slater determinant of the single particle states
\begin{equation}
    \Psi_{F}(x_{1},\cdots,x_{N})=\frac{1}{\sqrt{N!}}\det \left[\phi_{m}(x_{j})\right],
\end{equation}
where for the zero temperature ground state of the system the determinant only includes states whose $m$ and $j$ values range from $1$ to $N$.
Tonks-Girardeau gases have been experimentally observed \cite{kinoshita2004,paredes2004tonks} and are available today in different laboratories worldwide. 
Since this process allows one to obtain the full bosonic wave function, one has immediate access to the corresponding density and momentum distributions \cite{Pezer}. It is worth noting that all local properties, such as the density distribution, are identical for the TG gas and corresponding free Fermi gas, whereas properties that depend on the reduced single-particle density matrix, such as the momentum distribution, differ between them \cite{Minguzzi_tg,Pezer}. 

In the following, we will use the Tonks-Girardeau gas in the AFKP model as an accessible theoretical platform to study the effect of topological edge states on the Lieb excitations of a quantum many-body system. 

\section{The gapped Lieb excitation and topological flat mode}

Owing to the Bose--Fermi mapping theorem, the spectral function of a many-body TG gas can be exactly found using the single-particle wave functions \cite{Settino}.
For this one first calculates the lesser and the greater Green's function (see details in the Supplemental Material)
\begin{align}
    G^{<}(x,t;y,t^{\prime})&=i \left\langle \Psi^{\dagger}(x,t) \Psi(y,t^{\prime}) \right\rangle, \\
    G^{>}(x,t;y,t^{\prime})&=-i \left\langle \Psi(x,t) \Psi^{\dagger}(y,t^{\prime}) \right\rangle,
\end{align}
from which the Green functions in frequency-momentum space can be obtained via a Fourier transform as 
\begin{equation}
    G^{\gtrless}(\omega,k)=\int \! dx \, dy \, dt \, e^{-i(x-y)k}e^{i \omega t} G^{\gtrless}(x,t;y,t'=0).
\end{equation}
The spectral function in frequency-momentum space then follows as 
\begin{equation}
    A(\omega,k)=%-\frac{1}{\pi}\Im(G^{R}(\omega,k))=
    -\frac{1}{\pi}\Im\left(G^{>}(\omega,k)-G^{<}(\omega,k)\right),
\end{equation}
and provides an exact and general method to study how the many-body effects and the presence of single-particle edge states affect the excitation properties in strongly correlated many-body quantum systems. 
For simplicity we will choose in the following the particle number to be the same as the number of barriers, $N=M$, which means that at zero temperature the corresponding fermionic system fills all single particle states up to the first edge state. Furthermore, this sets the density as $\rho=N/L=1$ in our units.

\begin{figure*}[tb]
    \includegraphics[width=\linewidth]{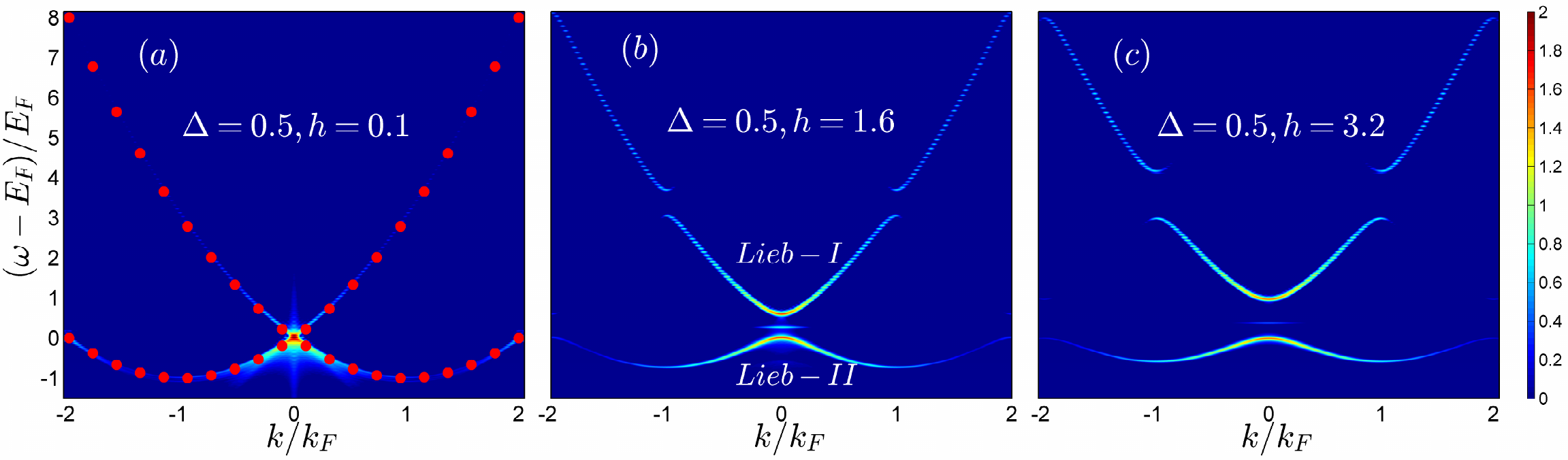} 
    \caption{Logarithmic spectral function of a Tonks-Girardeau gas in the AFKP model for different lattice heights (a) $h=0.1$, (b) $h=1.6$, and (c) $h=3.2$.  Note that the axes are scaled by $E_{F}=k_{F}^2/2$ and $k_{F}=\pi \rho$, with the density given by $\rho=N/L=1$. The other parameters are $\Delta=0.5$,  $L=40$, $M=40$, and  $N=40$. The red dots  in panel (a) correspond to the dispersion of the free TG gas.} 
    \label{fig:AFKPSpectrum}
\end{figure*}

The low energy many-body excitation behaviour of the AFKP model for different barrier heights is shown in Fig.~\ref{fig:AFKPSpectrum}. If the barrier height is very small (see panel (a)) the spectrum prominently features the standard Lieb-I (particle-like, going towards infinity) and Lieb-II (hole-like, excitation energy limited to within $E_{F}$) excitations of the free space Lieb-Liniger (LL) model.
In this limit the spectral profile also strongly overlaps with the dispersion of the free TG gas, given by $\epsilon(k)=k^{2}/2 \pm \pi \rho k$ \cite{Girardeau,Lieb2}, and indicated by red dots.
These two Lieb modes also remain visible for larger barrier heights (panels (b) and (c)), but several additional features stemming from the KP lattice and the edge mode start to emerge.

\begin{figure}[tb]
\centering
    \subfigure{ \includegraphics[scale=0.38]{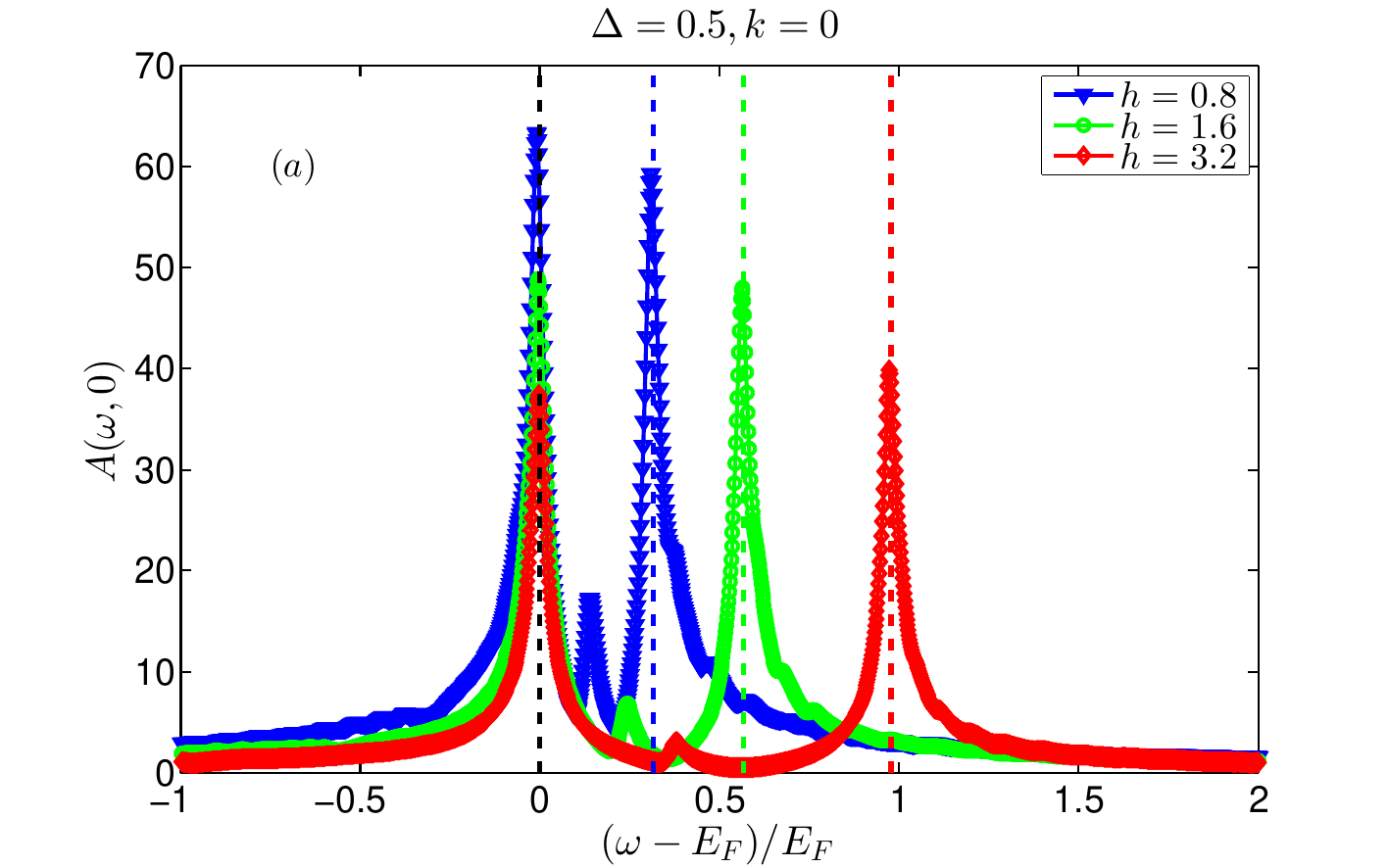} }
    \subfigure{  \includegraphics[scale=0.38]{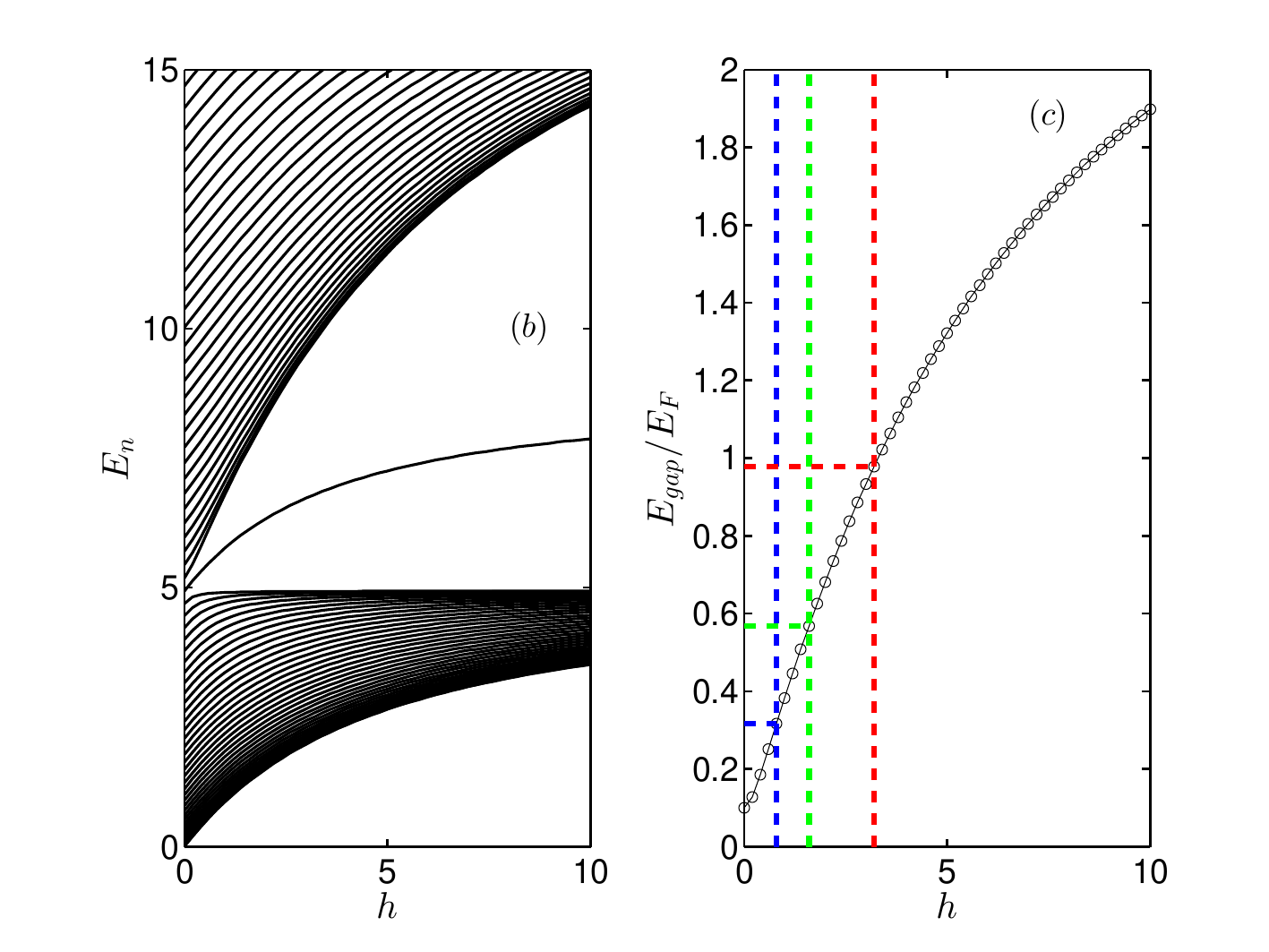} }
    \caption{ (a) Spectral function at $k=0$ for the barrier heights $h=[0.8,1.6,3.2]$ for a system with the same parameters as in Fig.~\ref{fig:AFKPSpectrum}. (b) The single particle energy spectrum of the AFKP model, and (c) the size of the gap in the single particle spectrum as a function of the barrier heights.}  
\label{fig_sf_line}
\end{figure}

The first is the opening of a gap between the two excitation branches at $(\omega,k)=(0,0)$. In the standard, free space LL model the two branches touch at this point, indicating that the energetic cost for particle creation and annihilation at $(\omega,k)=(0,0)$ is symmetric. 
The origin of this gap can be attributed to the lattice potential and it is directly related to the gap appearing in the single particle spectrum. One can immediately note that the gap width increases with increasing barrier height (see panels \ref{fig:AFKPSpectrum}(b) and \ref{fig:AFKPSpectrum}(c)), and to quantify this better we show in Fig.~\ref{fig_sf_line}(a) the spectral function at $k=0$ in detail for three different barrier heights. 
To confirm the consistency with the behaviour of the single particle energy spectrum, we show the latter in Fig.~\ref{fig_sf_line}(b) and the corresponding gap sizes in panel \ref{fig_sf_line}(c). Note that the gap sizes are defined as $E_{gap}=E_{M+1}-E_{M-1}$, as the $M^{th}$ state corresponds to the edge state.
The vertical dashed lines in panel (c) correspond to the three heights for which the cuts are shown in panel (a), and the corresponding single particle gap sizes are marked in panel (a) by dashed lines. An almost perfect match with the results obtained from the many-body SF calculation can be seen, confirming the direct relation between the many-body excitations and the single particle spectrum.

Next we turn our attention to the flat mode visible inside the gap of the spectral function. 
For the single particle spectrum the topological character of the edge state can be confirmed by calculating the corresponding Chern numbers \cite{Reshodko_2019} and within the first gap the system has maximally localized chiral edge states around $\Delta\approx\pm 0.5$. For $\Delta=\{-1,0,1 \}$ the edge state joins the respective lower and upper bands and its wavefunction becomes extended. This behaviour is also reflected in the many-body spectral function, as the wavefunction and energy of the edge mode play an important role in the elementary many-body excitations. 

\begin{figure*}
    \includegraphics[width=\linewidth]{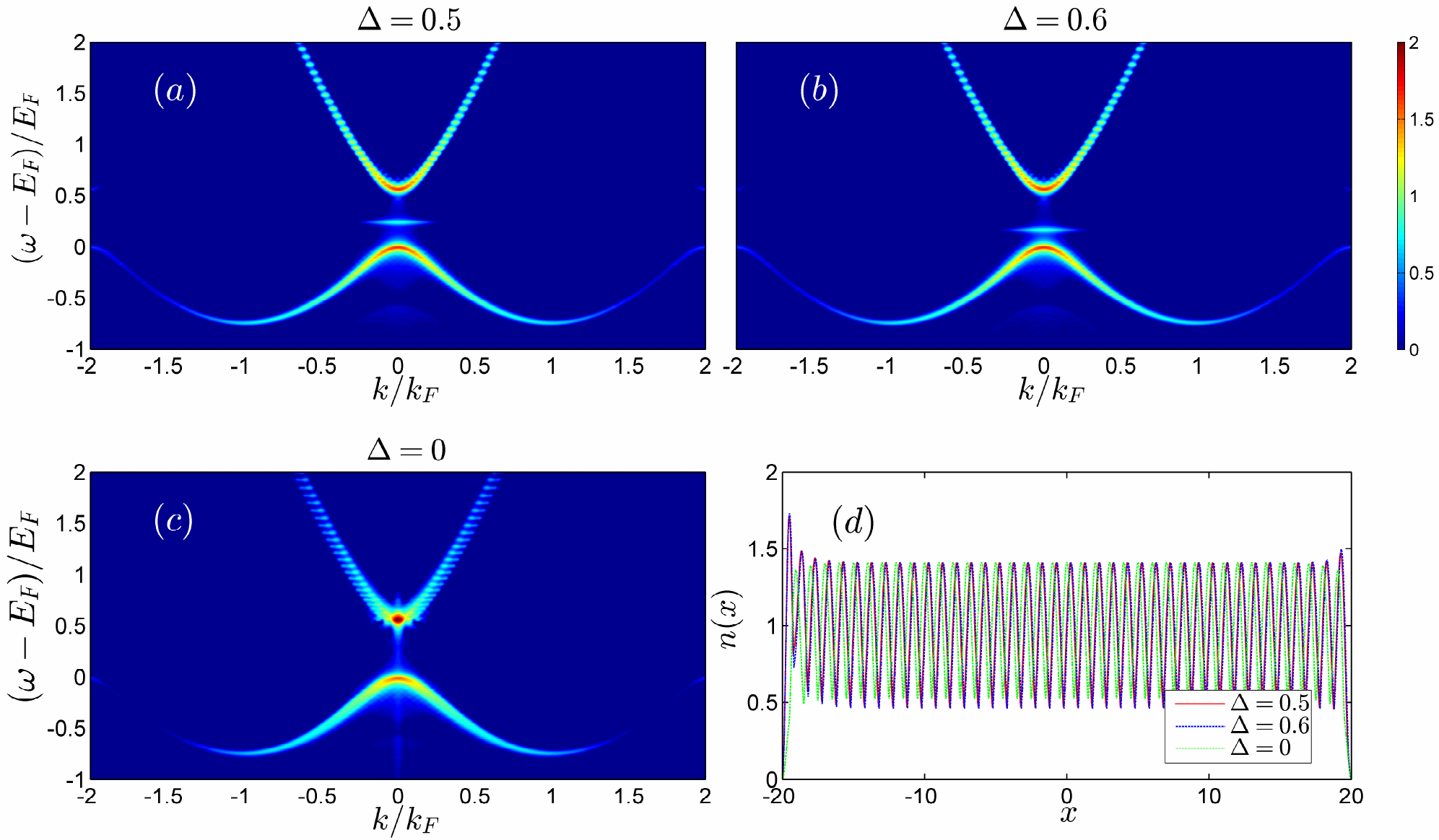}
    \caption{Logarithmic spectral function for different shift parameters (a) $\Delta=0.5$, (b) $\Delta=0.6$ and (c) $\Delta=0$. (d) The many-body density distribution of the ground state for the same three shift parameters. Here the barrier height is given by $h=1.6$ and all other parameters are the same as in Fig.~\ref{fig:AFKPSpectrum}. }
    \label{fig_GF_D}
\end{figure*}

To explore this we show in Figs.~\ref{fig_GF_D}(a)-(c) the spectral function for different values of the barrier shift. Looking at the situation with $\Delta=0.5$, for which a strongly localised single particle chiral edge state exists (see panels (a) and (d)), a strong flat mode is clearly visible around $k=0$ in the middle of the gap between the Lieb-I and the Lieb-II branches. For a barrier shift corresponding to $\Delta=0.6$ (panel (b)) the flat mode can be seen to move towards the Lieb-II band, whereas for $\Delta=0$ (panel (c)), the flat mode is absent and has merged with the Lieb-I branch. This is the case where all single particle eigenstates are fully delocalised and the spectral function should no longer feature any effects from the edge states. It is well worth noting that the gap between the two Lieb excitations remains unchanged under changes of $\Delta$, which again confirms that this is only controlled by the height of the periodic structure for fixed lattice constant.

\begin{figure}[tb]
    \centering
    \subfigure{\includegraphics[width=\linewidth]{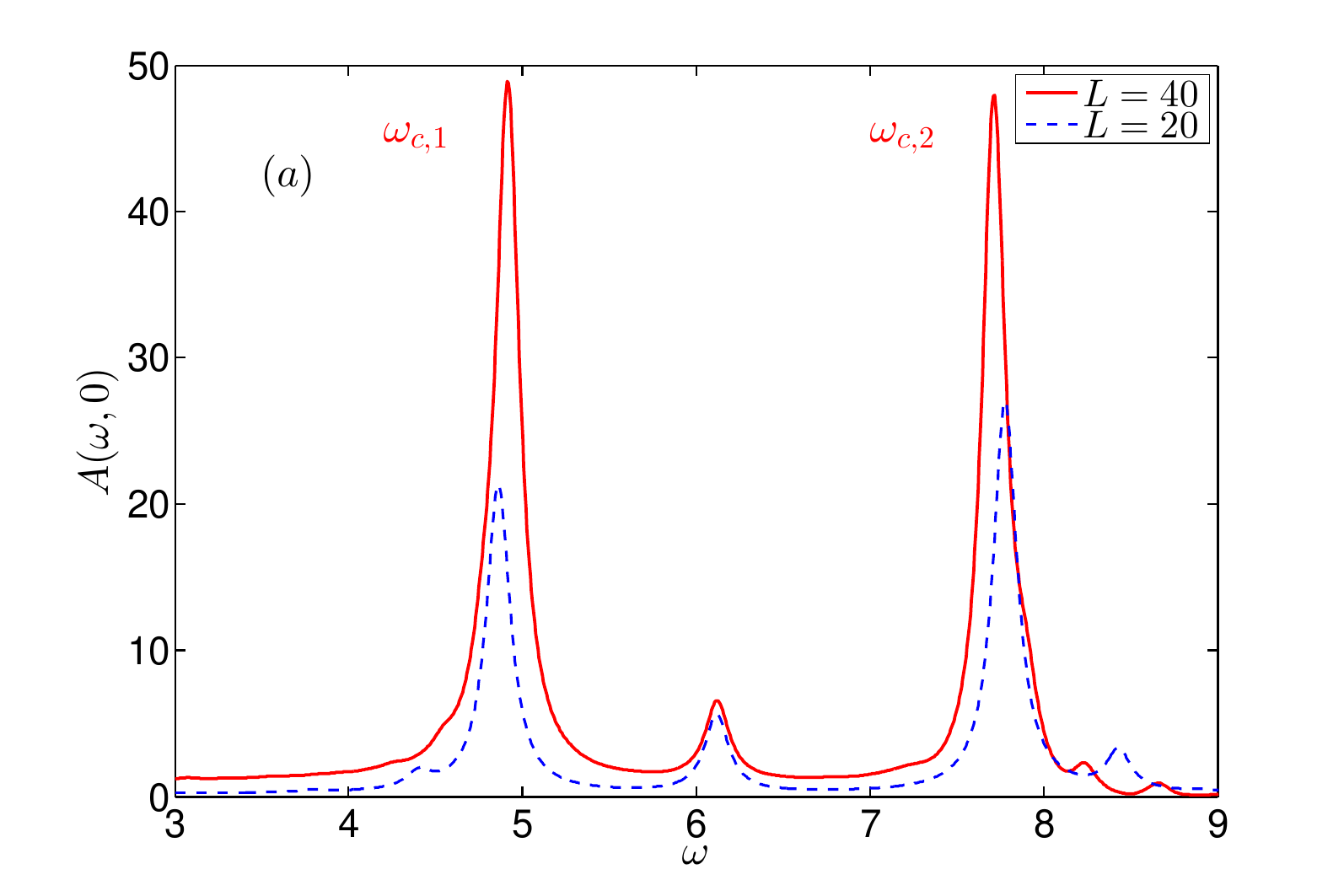} }
% {GF_ksin_LG_L40M40H1d6_N40_kw_NL256_scale_kwpi0.pdf
    \subfigure{\includegraphics[width=\linewidth]{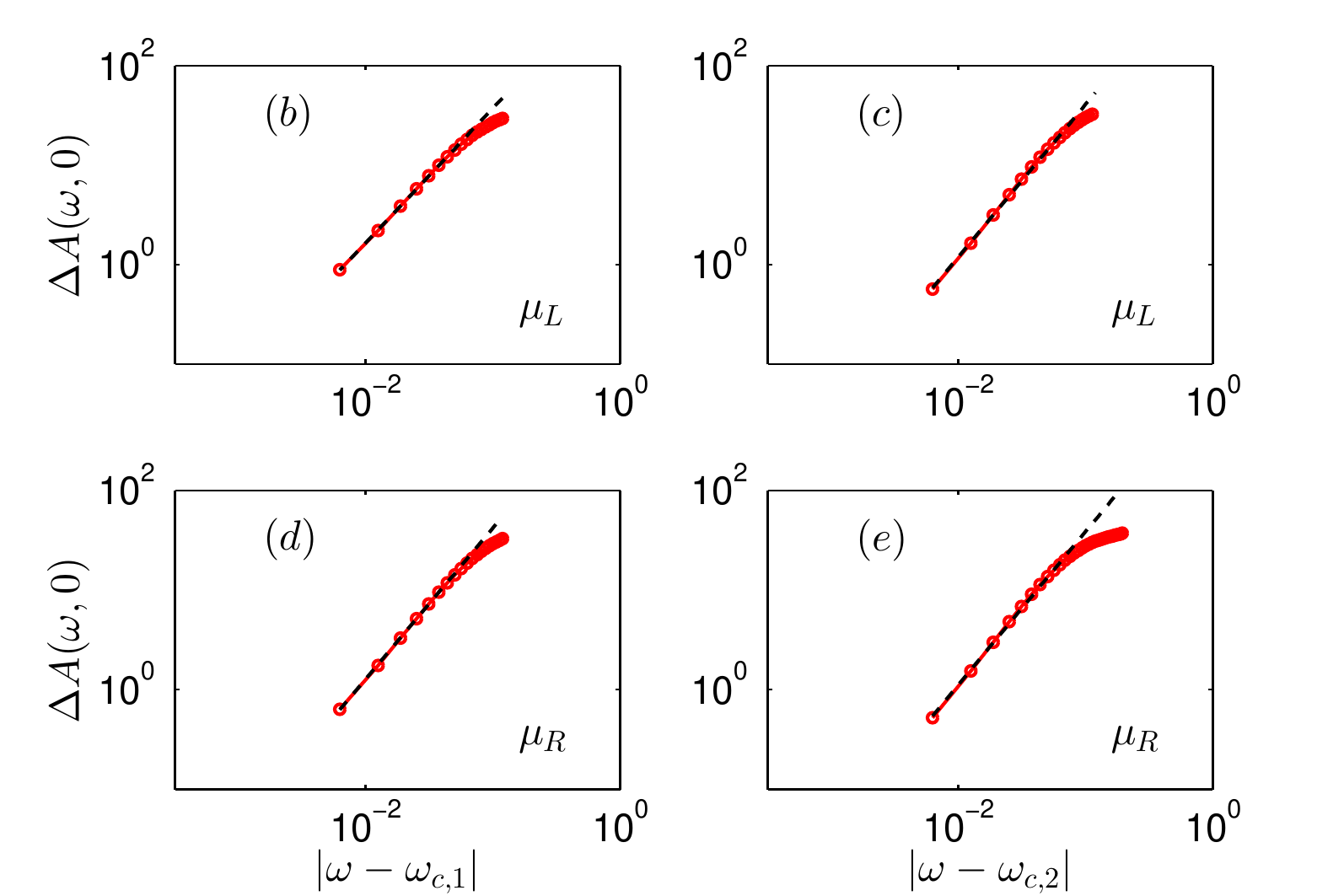}}
    \caption{(a) Spectral function at $k=0$ for a system with $\Delta=0.5$ and $ h=1.6$ for two sizes given by $L=M=N=[20,40]$. (b)-(e) The slope of the two singular peaks as a function of the frequency in logarithmic scale for $L=40$.  The dashed black lines are obtained by numerical fitting and panels (b) and (c) correspond to $\mu_{L}\sim [1.35,1.53]$, while panels (d) and (e) correspond to $\mu_{R}\sim [1.51,1.53]$. } 
    \label{fig_sf_scale}
\end{figure}

In Fig.~\ref{fig_sf_scale}(a) we show the spectral function at $k=0$ for two system sizes, $L=20$ and $L=40$.   Here again the left most and right most peaks correspond to the Lieb-II and Lieb-I excitations, respectively, and their divergence is reminiscent of effect of the Fermi edge singularities (FES) in condensed matter systems \cite{Glazman2007,Glazman2008,gogolin_book}. This effect is well studied in strongly interacting one-dimensional electronic systems and corresponds to an enhancement of spectral function near the Fermi point \cite{Ogawa}.
However, even though the TG gas we are considering is bosonic and its spectral function is different from the one for a Fermi gas (see Fig.~\ref{fermi_L40M40H1d6_N40_kw_NL256_D0d5} in the Supplemental Material), it is natural to expect the FES to also appear here due to the one-dimensional geometry and the infinite interactions of the model.

 An analysis of the divergence for the two Lieb excitations peaks for the case of $L=40$ is shown in panels \ref{fig_sf_scale}(b)-\ref{fig_sf_scale}(e). In order quantify the diverging behaviour,  we calculate the slope of the left and the right side of each peak along the $\omega$-axis, and fit a power-law scaling relation for the spectral function of the form\cite{Glazman2008,gogolin_book} 
\begin{equation}
\Delta A(\omega,k=0) \sim |\omega-\omega_{c}|^{-\mu_{L,R}},
\end{equation}
where $\mu_{L}$ quantifies the scaling on the left and $\mu_{R}$ on the right hand side of the peak, while $\omega_{c}$ represents the position of the peak of the Lieb excitations. One can see that numerically fitting to the power law behaviour gives good results, and the exponents of Lieb excitations in this case are found to be close to $1.5$.  
Contrary to the behaviour of the two Lieb modes, the peak corresponding to the topological edge mode does not significantly change with the system length and particle number. This indicates the absence of a singularity in the spectral around the flat edge mode, which can be understood by noting that the edge state is separated from the low and upper bands by a gap. The density of states near the edge state is therefore  small compared to close to the edges of the bands above and below it, leading to less opportunities for low-energy excitation. 

\begin{figure*}
    \includegraphics[scale=0.5]{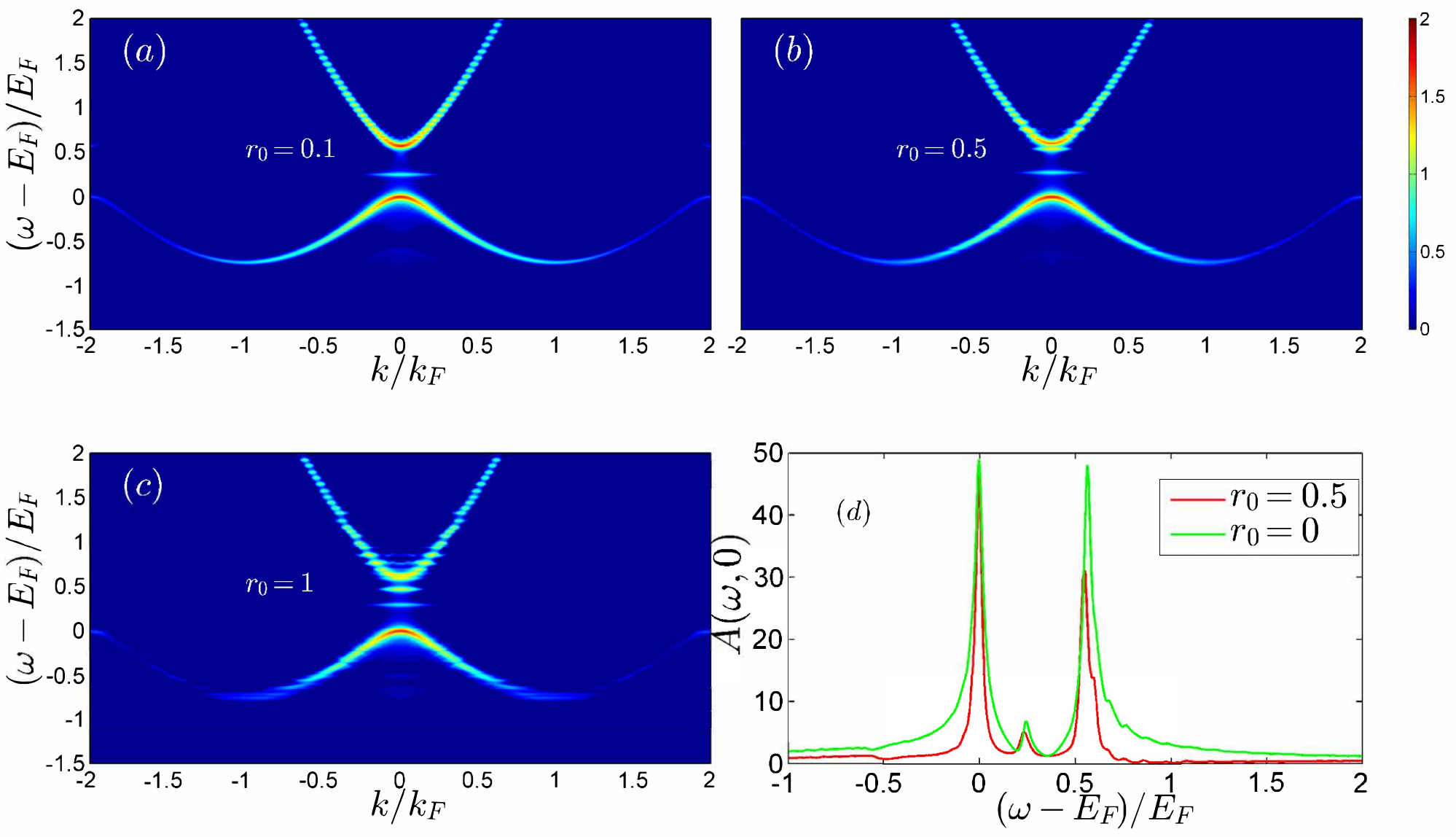}
    \caption{ (a)-(c) Logarithmic spectral function for a system with randomly perturbed barrier heights $h_{i}=\bar{h} +\delta h_{i}$ with amplitudes $\delta h_{i} \in [-r_0,r_0]$ and $\bar{h}=1.6$. The values of the respective $r_0$ are indicated in the individual panels. Here again $\Delta=0.5, L=40, M=40$ and $N=40$. (d) Cut of the of spectral function along $k=0$. The red curve is averaged over ten different realisations for $r_0=0.5$, while the green curve corresponds to the unperturbed situation. }
    \label{GF_rand_L40M40}
\end{figure*}

Let us finally confirm the robustness of edge mode in the spectral function. For this we consider a system with perturbations in the barrier heights, $h_{i}=h+\delta h_{i}$, with $\delta h_{i}$ randomly chosen from the interval $[-r_0,r_0]$. Examples of the spectral function for three perturbation amplitudes $r_0$ for a mean height of $h=1.6$ are shown in Fig.~\ref{GF_rand_L40M40}(a)-(c). One can see that, as expected, the Lieb excitations become more blurred by the random perturbations, while the flat edge mode between two Lieb excitations remains mostly unchanged.  We also show a cut of the spectral function along $k=0$ averaged over ten different samples in Fig.~\ref{GF_rand_L40M40}(d) for the same value of the average height. One can clearly see that the two Lieb excitations are significantly affected, however the peak corresponding to the flat edge mode is mostly unaffected. This confirms that the flat edge mode is robust to random perturbation in the bulk, even in strongly correlated systems.

\section{Conclusion}
\label{Sec:Conclusion}

We have studied the spectral function of a strongly correlated  Tonks-Girardeau gas in a finite sized Kronig-Penney potential. Since the single particle eigenfunctions of this model can be found, the exact many-body spectral function can be constructed and we have discussed two key observations from it. First, the Lieb-I and Lieb-II excitations become gapped as a function of the barrier height and second, a robust topological flat mode appears in this gap near zero momentum. We have also established the divergent behaviour of the two Lieb modes, similar to effects found in corresponding fermionic systems, and shown that the gap modes is robust against perturbations in the system even in the presence of strong correlations. Our results contribute to the understanding of the many-body excitations in interacting quantum gases with non-trivial topological properties \cite{tanese2013} and can in principle be observed in systems of ultracold atom trapped in nano-scale potential lattices \cite{Zoller2016,Kruckenhauser,Subhankarprx}.

\section{acknowledgments}
We thank Profs.~G.~Juzeli\= unas, T.~Ozawa and D.~Blume for fruitful discussions. This work was supported by the Okinawa Institute of Science and Technology Graduate University and utilized the computing resources of the Scientific Computing and Data Analysis section of the Research Support Division at
OIST.

\bibliography{kpspec}

\onecolumngrid
\newpage
\appendix
\begin{center}
    \textbf{Appendix:  The Lieb  excitation and topological flat mode of  spectral function of Tonks-Girardeau gas in Kronig-Penney potential}
\end{center} 
%%Supplemental material
\begin{center}
Wen-Bin He, Giedrius \v{Z}labys, Hoshu Hiyane, Sarika Sasidharan-Nair, and Thomas Busch    
\end{center} 

In this appendix we provide more details and additional results that help understand and follow the main presentation.

\section{RSPDM for the many-body TG gas}

For a bosonic quantum many-body system consisting of $N$ particles, which is described by the wavefunction $\psi_{B}(x_1,x_2,\cdots,x_N)$, the reduced singe-particle density matrix (RSPDM) is defined as 
\begin{equation}
\rho_{B}(x,y)=N \int dx_{2} \cdots dx_{N} \psi_{B}^{*}(x,x_{2},\cdots,x_{N})\psi_{B}(y,x_{2},\cdots,x_{N}).
\end{equation}
While this expression is difficult to calculate in general, in the TG limit it can be rewritten in terms of the single particle eigenstates $\phi_{j}(x)$ as  \cite{Pezer} 
\begin{equation}
\rho_{B}(x,y)=\sum_{i,j=1}^{N} \phi_{i}^{*}(x) \mathbf{A}_{ij}(x,y) \phi_{j}(y),
\end{equation}
where the $N \times N $ matrix $\mathbf{A}(x,y)=(\mathbf{Q}^{-1})^{T} \det(\mathbf{Q})$, and the elements of $\mathbf{Q}$ are given by (assuming $x<y$)
\begin{equation}
Q_{ij}(x,y)=\delta_{ij}-2  \int_{x}^{y} du \phi_{i}^{*}(u)\phi_{j}(u).
\end{equation}
The diagonal elements of RSPDM correspond to the spatial density distribution of the TG gas, as for example shown in Fig.~\ref{fig_GF_D}(d). 

\section{Green's function of a many-body Tonks-Girardeau gas}
\noindent
In general the lesser and greater Green's functions are defined as 
\begin{align}
    G^{<}(x,t;y,t')&=i \left\langle \Psi^{\dagger}(x,t) \Psi(y,t') \right\rangle\;, \\
    G^{>}(x,t;y,t^{\prime})&=-i \left\langle \Psi(x,t) \Psi^{\dagger}(y,t') \right\rangle\;, 
\end{align}
where the $\Psi$ are the full many-body wavefunctions. As shown in \cite{Settino}, these can be rewritten for a Tonks-Girardeau gas in terms of the single particle eigenstates as
\begin{align}
    -iG^{<}(x,t;y,t')&=\det[\mathbf{P}(x,t) \mathbf{P}(y,t')|_{\eta}]a^{<}(x,t;y,t'),\\
    iG^{>}(x,t;y,t')&=\det[\mathbf{P}(x,t) \mathbf{P}(y,t')|_{\eta}]a^{>}(x,t;y,t'),
\end{align}
where the elements of the key matrix $\mathbf{P}$ are given as 
%\tb{are these the same matrix elements as in the section above? } \wb{ using different notation. they are different}
\begin{equation}
    P_{lm}(x,t)=\delta_{lm}-2e^{-it(\epsilon_{l}-\epsilon_{m})} \int_{x}du \phi_{l}(u)\phi^{*}_{m}(u)\;,
\end{equation}
with the single-particle wave functions given by $\phi(x,t)=\left[\phi_{1}(x,t),\cdots,\phi_{M}(x,t)\right]^T$. Furthermore
\begin{align}
    a^{<}(x,t;y,t')&=\phi^T(x,t)_{\eta} \left\{ [\mathbf{P}(x,t) \mathbf{P}(y,t')]^{-1}\right\}^{T}_{\eta} \phi^*(y,t')_{\eta},\\
    a^{>}(x,t;y,t')&=\phi^\dagger(y,t')\phi(x,t)- \left[\phi^\dagger(y,t') \mathbf{P}(x,t)\right]_{\eta} \left[\mathbf{P}(x,t) \mathbf{P}(y,t')\right]^{-1}|_{\eta} [\mathbf{P}(y,t')\phi(x,t)]_{\eta},
\end{align}
and the subscript ${\eta}$ indicates that the results should be projected onto first $N$ lowest energy states.  Finally, the spectral function is given in terms of the retarded  Green's function and therefore the lesser and greater Green's function as
\begin{equation}
    A=-\frac{1}{\pi}\Im(G^{R})=-\frac{1}{\pi}\Im(G^{>}-G^{<}).
\end{equation}
The above formalism only requires the knowledge of the single-particle wave functions to obtain the spectral function of  many-body system, regardless of the trapping potential of the TG gas.

\section{Higher Edge Modes}
\noindent

\begin{figure}[tb]
    \includegraphics[width=\linewidth]{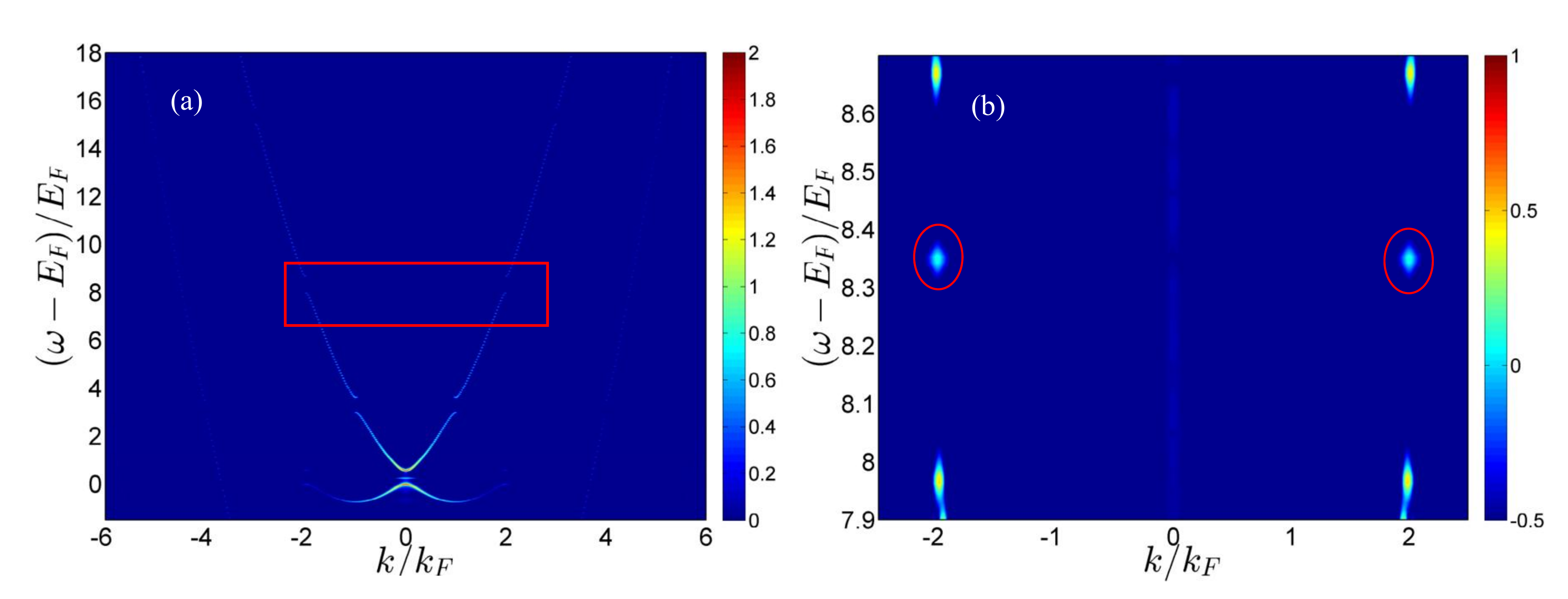} %%% kp_sf_h_hband_v3
    \caption{(a) Logarithmic spectral function of a Tonks-Girardeau gas in the AFKP model for higher energies for a lattice of height $h=1.6$. (b) Zoom into the third gap. Note the different color scale. Here $\Delta=0.5$,  $L=40$ and  $M=40$.  } 
    \label{h1d6_hband_zoom}
\end{figure}

It is worth noting that the TG model is not only a low energy theory, but that it applies to the excitations of a strongly correlated gas at higher energies as well. Since edge states also exist in the higher lying gaps of the single particle spectrum, one can therefore expect them to appear at higher energies in the spectral function. To confirm this, we show in  Fig.~\ref{h1d6_hband_zoom}(a) the spectral function for higher energies for a barrier height of $h=1.6$, in which gaps at higher lying energies are also clearly visible. To check for the presence of edge states in the gaps, we show in panel (b) 
a zoom into the third gap where the presence of the edge states is clearly visible. 

However, for the chosen value of $\Delta=0.5$, the edge modes in the second gap of the single particle spectrum are very close to the next higher lying band, which means that in the spectral function they do not visibly disconnect from the next higher part of the spectral branch within the resolution we can achieve.

\section{Fermionic Spectral Function}
\noindent

\begin{figure}[tb]
    \includegraphics[width=0.6\linewidth]{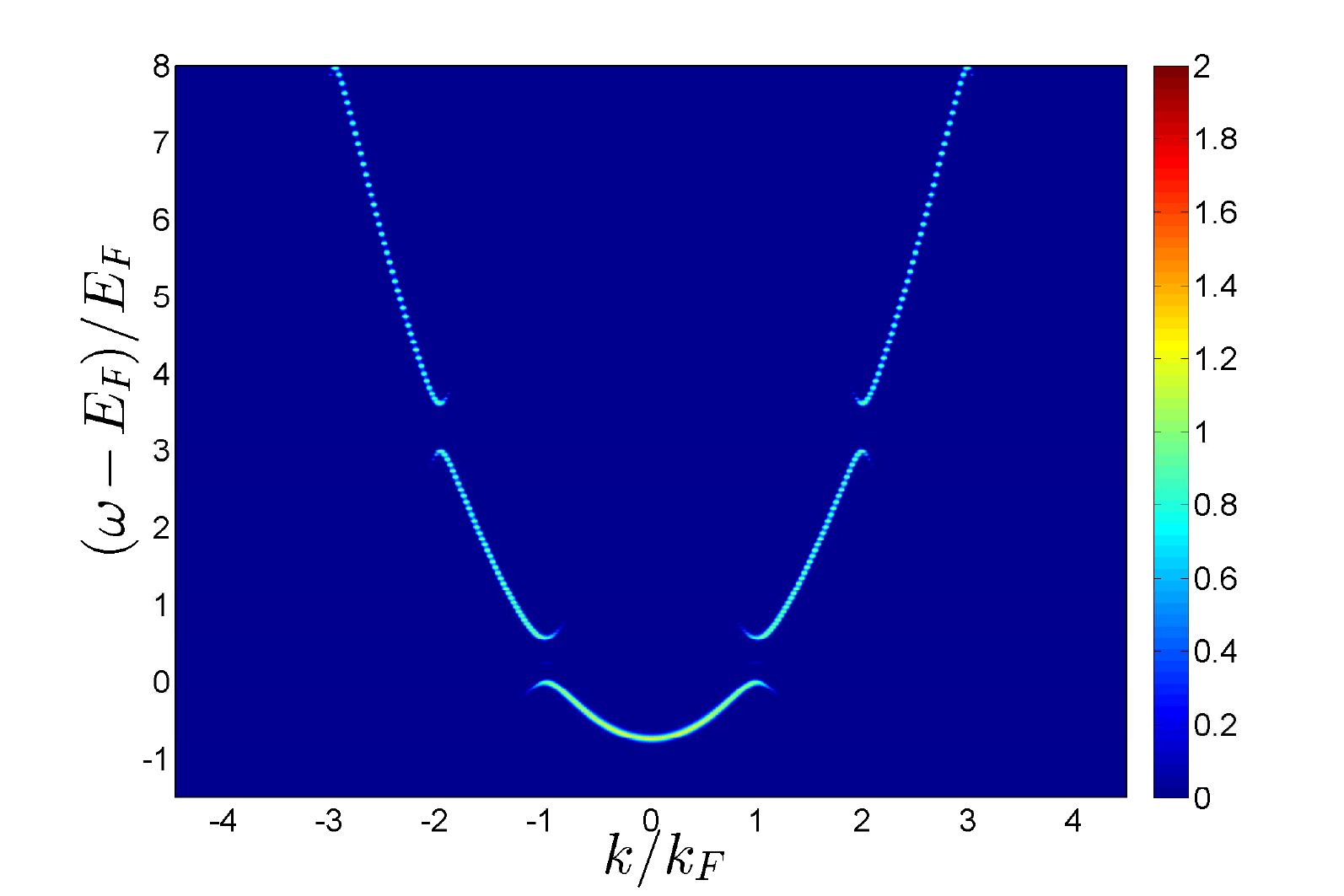}
    \caption{Logarithmic spectral function of a free Fermi gas in the AFKP. The parameters here are $\Delta=0.5$, $L=40, M=40, N=40$ and $h=1.6$.}
\label{fermi_L40M40H1d6_N40_kw_NL256_D0d5}
\end{figure}

Since the non-local properties of a bosonic TG and a free fermion gas are known to be different, it is interesting to briefly also look at the spectral function of a Fermi gas.
One can immediately notice fundamental differences from Fig.~\ref{fermi_L40M40H1d6_N40_kw_NL256_D0d5} (compare to the TG case shown in Fig.~\ref{fig:AFKPSpectrum}(b)), which are expected due to the different single particle momentum  distributions. For the fermions the gap and the edge mode are located around the Fermi momentum $k/k_{F}=1$, and not around $k/k_{F}=0$ as for the bosons. Furthermore, each spectral band has an almost constant excitation weight, which is due to the form of the Fermi distribution at zero temperature. 

\end{document}